\documentstyle[12pt]{article}
\begin{document}
\title{\bf{Quantum mechanics over a q-deformed
\\ (0+1)-dimensional superspace}}
\author{\bf{H. Montani and R. Trinchero}
\\{\normalsize \it{Centro At\'{o}mico Bariloche and Instituto Balseiro}}
\\{\normalsize \it{\ 8400 - S. C. de Bariloche, Rio Negro,}}
\\{\normalsize \it{Argentina.}}
}
\date{}
\maketitle
\begin{abstract}
We built up a explicit realization of (0+1)-dimensional q-deformed
superspace coordinates as operators on standard superspace. A
q-generalization of supersymmetric transformations is obtained,
enabling us to introduce scalar superfields and a q-supersymmetric
action. We consider a functional integral based on this action.
Integration is implemented, at the level of the coordinates and
at the level of the fields, as traces over the corresponding
representation spaces. Evaluation of these traces lead us to standard
functional integrals. The generation of a mass term for the fermion
field leads, at this level, to an explicitely broken version of
supersymmetric quantum mechanics.
\end{abstract}

\newpage
In the last few years the idea that non-commutative geometry could play a
central role in the formulation of fundamental physics has attracted some
attention\cite{con}. The Connes-Lott version of the Standard
Model\cite{conl} and Quantum Groups, regarded as endomorphism of
some non-conmutative space\cite{man}, are important examples supporting this
idea.
In this paper, by means of a simple example, we address the
question, how to do quantum mechanics over a non-commutative space?.  Among
the non-commutative spaces, the quantum plane is an example strongly related
to some physical systems. It is clearly understood and, furthermore, the
endomorphism of its algebra leads to Quantum Groups\cite{drinf}\cite{man},
an structure underlying the integrability of physical models\cite{fad}.  The
quantum plane involves two homogeneously non-commutative "bosonic"
coordinates.In this paper we consider an even simpler example well adapted
for our purposes. We deal with only one "bosonic" coordinate and two
fermionic ones sharing properties with Grassmann variables. As in the
quantum plane these coordinates are assumed to be homogeneously
non-commutative. We represent the resulting algebra as operators over a
standard $(0+1)$-dimensional superspace. Let $t,\eta,{\bar \eta}$ be the
coordinates in this standard superspace. The ${\bar .}$ operation can be
considered as an involution and we assign a Grassmann number $\sharp$ to
each coordinate, i.e., \begin{eqnarray} \begin{array}{cc} t\longrightarrow
\stackrel{\_}{(t)}=t & \eta \longrightarrow \overline{(\eta
)}=\overline{\eta }\nonumber\\ \sharp t=0 & \sharp \eta =1 =- \sharp
\overline{\eta }\;, \end{array}
\label{inv} \end{eqnarray}
the algebraic relations between the coordinates are, \begin{equation}
\begin{array}{l} \lbrack t,\eta ]^{^{-}}\equiv t\eta -\eta t=0=
\lbrack \stackrel{\_}{\eta },t]^{^{-}}\equiv \overline{\eta }t-t
\overline{\eta } \\
\\
\lbrack \overline{\eta },\eta ]^{^{+}}\equiv \overline{\eta }%
\eta +\eta \overline{\eta }=0=\lbrack \eta ,\eta ]^{^{+}}
=[\overline{\eta },\overline{\eta }%
]^{^{+}}
\end{array}
\label{ssalg}
\end{equation}
Now, let us consider the following operator valued fields on the superspace
$\{t,\eta ,\overline{\eta }\}$,
\begin{equation}
\begin{array}{l}
\theta =\eta e^{-\frac 12\alpha \overline{\eta }\partial _{\overline{\eta }%
}}e^{i\alpha tT}=\eta [1+(e^{-\alpha /2}-1)\overline{\eta }\partial _{%
\overline{\eta }}]e^{i\alpha tT} \\
\\
\overline{\theta }=\sqrt{q}e^{-i\alpha tT}e^{\frac 12\alpha \eta \partial
_\eta }\overline{\eta }=\sqrt{q}[1+(e^{\alpha /2}-1)\eta \partial _\eta ]%
\overline{\eta }e^{-i\alpha tT}
\end{array}
\label{teta}
\end{equation}
where $\alpha=\ln q$ , $q\in R\!\!\!\!R$ and $T$ is the Hermitian
generator of time translations satisfying $[T,t]=i$. The set $\{t,\theta ,%
\overline{\theta }\}$ has the following commutation relations,
\begin{equation}
\begin{array}{l}
\lbrack t,\theta ]_q^{^{-}}\equiv t\theta -q\theta t=0=
\lbrack \stackrel{\_}{\theta },t]_q^{^{-}}\equiv \overline{\theta }t-qt%
\overline{\theta } \\
\\
\lbrack \stackrel{\_}{\theta },\theta ]_q^{^{+}}\equiv \overline{\theta }%
\theta +q\theta \overline{\theta }=0=
\lbrack \theta ,\theta ]^{^{+}}=[\overline{\theta },\overline{\theta }%
]^{^{+}}
\end{array} \;.
\label{qssalg}
\end{equation}
Moreover, this algebra is preserved under the q-deformed supersymmetric
transformations,
\begin{equation}
\begin{array}{lll}
\delta _\varepsilon t=-i\overline{\theta }\varepsilon &  & \delta _{%
\overline{\varepsilon }}t=i\overline{\varepsilon }\theta \\
\delta _\varepsilon \theta =\varepsilon &  & \delta _{\overline{\varepsilon }%
}\theta =0 \\
\delta _\varepsilon \overline{\theta }=0 &  & \delta _{\overline{\varepsilon
}}\overline{\theta }=\overline{\varepsilon }
\end{array}
\label{qsstrans} \;,
\end{equation}
with the parameters $\varepsilon $ and $\overline{\varepsilon }$ satisfying,
\begin{equation}
\begin{array}{l}
\lbrack t,\varepsilon ]_q^{^{-}}=[\overline{\varepsilon },t]_q^{^{-}}=0 \\
\lbrack \theta ,\varepsilon ]_q^{^{+}}=[\stackrel{\_}{\theta },\overline{%
\varepsilon }]_q^{^{+}}=0 \\
\lbrack \overline{\varepsilon },\theta ]_q^{^{+}}=[\stackrel{\_}{\theta }%
,\varepsilon ]_q^{^{+}}=0 \\
\lbrack \overline{\varepsilon },\varepsilon ]_q^{^{+}}=0
\end{array}
\label{tetapar}
\end{equation}
and the fundamental supersymmetric relation,
\begin{equation}
\lbrack \delta _\varepsilon ,\delta _{\overline{\varepsilon }%
}]^{-}=2i\varepsilon \overline{\varepsilon }\partial _t  \;.
\label{qsusy}
\end{equation}
The parameters $\varepsilon $ and $\overline{\varepsilon }$ can be expressed
also as operators over the superspace $\{t,\eta ,\stackrel{
\_}{\eta }\}$ as
\begin{equation}
\begin{array}{l}
\varepsilon =\zeta e^{-\frac 12\alpha \stackrel{\_}{\eta }\partial _{%
\stackrel{\_}{\eta }}}e^{-\alpha \stackrel{\_}{\zeta }\partial _{\stackrel{\_%
}{\zeta }}}e^{i\alpha tT} \\
\\
\stackrel{\_}{\varepsilon }=qe^{-i\alpha tT}e^{-\alpha \zeta \partial _\zeta
}e^{\frac 12\alpha \eta \partial _\eta }\stackrel{\_}{\zeta }
\end{array} \;.
\label{epsilon}
\end{equation}
As usual, one may now introduce a scalar $q$-superfield $\Phi$,
\begin{equation}
\Phi (t)=x(t)+i\theta \psi (t)-i\overline{\psi }(t)\overline{\theta }+%
\overline{\theta }\theta d(t)  \label{qssfield}
\end{equation}
The scalar transformation property of this field under (\ref{qsstrans}),
induce the following transformation rules for the
$q$-superfield components $\{x(t),\psi (t),\overline{\psi }(t),d(t)\}$
\begin{equation}
\begin{array}{l}
\delta _qx(t)=-i\varepsilon \psi (t)+i\overline{\psi }(t)\stackrel{\_}{%
\varepsilon } \\
\\
\delta _q\psi (t)=-iq\overline{\varepsilon }\{[T,x(t)]^{^{-}}-d(t)\} \\
\\
\delta _q\overline{\psi }(t)=-iq\{[T,x(t)]^{^{-}}+d(t)]\varepsilon  \\
\\
\delta _qd(t)=-{i\over q}\{\varepsilon [T,\psi (t)]_q^{^{-}}+[\overline{\psi }%
(t),T]_q^{^{-}}\overline{\varepsilon }\}
\end{array}\;.
\label{fsstrans}
\end{equation}
There  commutation relations
between $\{t,\theta ,\stackrel{\_}{\theta }\}$ and the $q$-superfield
components $\{x(t),\psi (t),\overline{\psi }(t),d(t)\}$ are non trivial.
Regarding $x(t)$ as a power series in the operator $t$ with complex
coefficients, and requiring the invariance of the commutation relations
under the $q$-supersymmetric transformations (\ref{qsstrans}), we get,
\begin{equation}
\begin{array}{lll}
x(t)\theta =\theta x(qt) &  & x(t)\overline{\theta }=\overline{\theta }x(t/q)
\\
\\
\psi (t)\theta =-\theta \psi (qt) &  & \psi (t)\overline{\theta }=-q%
\overline{\theta }\psi (t/q) \\
\\
\overline{\psi }(t)\theta =-q^{-1}\theta \overline{\psi }(qt) &  & \overline{%
\psi }(t)\overline{\theta }=-\overline{\theta }\overline{\psi }(t/q) \\
\\
x(qt)\psi (t)=\psi (t)x(t) &  & x(t)\overline{\psi }(t)=\overline{\psi }%
(t)x(qt) \\
\\
d(t)\theta =\theta d(qt) &  & d(t)\overline{\theta }=q^{-1}\overline{\theta }%
d(t/q)
\end{array}
\label{10}
\end{equation}
\[
\psi (t)\overline{\psi }(t)=-q^{-1}\overline{\psi }(qt)\psi (qt) \;.\]

It is worth remarking that the requirement of $q$-supersymmetric invariance
for the relations in the second line of (\ref{10}) involves objects such as
$\delta \psi (qt)$. Such variations
should be obtained from the variation of the superfield
$\Phi (qt)$ under the transformation $\delta _q(qt)=q\delta _qt=qi(\overline{%
\varepsilon }\theta -\overline{\theta }\varepsilon ),$ for which
the time evolution operator is now $T^{\prime }=T/q$ ,
satisfying $[T^{\prime },qt]=i.$

The Grassmannian fields $\psi (t)$ and $\overline{\psi }(t)$ can be
represented as,
\begin{equation}
\begin{array}{l}
\psi (t)=e^{-i\alpha tT}e^{\frac 12\alpha \eta \partial _\eta }\chi (t) \\
\\
\overline{\psi }(t)=q^{-1/2}\overline{\chi }(t)e^{-\frac 12\alpha \overline{%
\eta }\partial _{\overline{\eta }}}e^{i\alpha tT}
\end{array}
\label{psi}
\end{equation}
where $\chi (t)$ and $\stackrel{\_}{\chi }(t)$ are $q$-Grassmannian
fields satisfying the algebra,
\begin{equation}
\begin{array}{c}
\chi (t)\overline{\chi }(t)=-q^{-2}\overline{\chi }(t)\chi (t) \\
\\
\chi (t)^2=0=
\overline{\chi }(t)^2
\end{array}
\label{xi}
\end{equation}
and the following (anti)commutation relation with the usual superspace
coordinates,
\begin{equation}
\begin{array}{ccc}
\lbrack \chi (t),t]^{^{-}}=0 & , & [\stackrel{\_}{\chi }(t),t]^{^{-}}=0 \\
\\
\lbrack \chi (t),\eta ]^{^{+}}=0 & , & [\stackrel{\_}{\chi }(t),\overline{%
\eta }]^{^{+}}=0 \\
\\
\lbrack \chi (t),\partial _\eta ]^{^{+}}=0 & , & [\stackrel{\_}{\chi }%
(t),\partial _{\overline{\eta }}]^{^{+}}=0
\end{array}
\label{xitn}
\end{equation}
Now, we introduce the covariant derivatives,
\begin{equation}
\begin{array}{l}
D_\theta \equiv \partial _\theta -q\overline{\theta }T \\
\\
D_{\overline{\theta }}\equiv \partial _{\overline{\theta }}-qT\theta
\end{array}
\label{covder}
\end{equation}
which allows us to build up a $q$-supersymmetric invariant Lagrangian,
\begin{equation}
L=\frac 12D_{\overline{\theta }}\Phi D_\theta \Phi -V(\Phi )  \label{sslagr}
\end{equation}
where $V(\Phi )$ is some polynomial function.

Defining an action on the
superspace $\{t,\eta ,\overline{\eta }\}$ requires the analogous of
integration of $L$ over $t$, $\eta $ and $\stackrel{\_}{\eta }$. For
the case of our operator valued Lagrangian this can be achieved
by taking traces as a replacement
of integration. Therefore we need to have an inner product
defined on our function space. For the $t$ dependence we take
the usual one of square integrable functions, for the Grassmannian
sector we use,
\begin{equation}
<f\mid g>\equiv \int d\eta d\overline{\eta }\; \overline{f(\eta ,%
\overline{\eta })}\,g(\eta ,\overline{\eta })  \label{inpro}
\end{equation}
where the standard Berezin rules for integration on Grassmannian
variables are assumed. Hence we define the action by,
\begin{equation}
S(\Phi )=\{Tr_{\{t,\eta ,\overline{\eta }\}}[L(\Phi )]\}
=\int_0^Tdt
\; \sum_n <t,e_n \mid L(\Phi )\mid e_n,t>  \;,
\label{accion}
\end{equation}
where  $\{e_n\}$ denotes a basis for the space of functions
of $\eta$ and${\overline {\eta}}$. If we choose that basis to be,
\begin{equation}
e_0 ={1\over{\sqrt{2}}}(1 +{\overline{\eta }}\eta ) \;,e_1=\eta \;
,e_2=\overline{\eta } \;,
e_3={1\over{\sqrt{2}}}(1-{\overline{\eta }}\eta )\;\;,
\label{basis}
\end{equation}
its matrix of inner products is,
\begin{equation}
<e_i\mid e_j>=\left[
\begin{array}{llll}
1 & 0 & 0 & 0 \\
0 & 1 & 0 & 0 \\
0 & 0 & -1 & 0 \\
0 & 0 & 0 & -1
\end{array}
\right]\;,  \label{metric}
\end{equation}
This representation space have negative norm
states. However, this pseudo-metric leads to an Hermitian inner product,
\begin{equation}
<f\mid g>^{*}=<g\mid f>  \;,
\label{hemitian}
\end{equation}
We employ the following results in the evalation of traces,
\begin{equation}
\begin{array}{lllllll}
\eta e_0={{e_1}\over{\sqrt{2}}} &  & \overline{\eta }%
e_0={{e_2}\over{\sqrt{2}}} &  & \partial _\eta
e_0={{-e_2}\over{\sqrt{2}}} &  & \partial _{\overline{%
\eta }}e_0={{e_1}\over{\sqrt{2}}} \\
&  &  &  &  &  &  \\
\eta e_1=0 &  & \overline{\eta }e_1={{(e_0
-e_3)}\over{\sqrt{2}}} &  & \partial _\eta e_1={{%
(e_0+e_3)}\over{\sqrt{2}}} &  & \partial _{\overline{\eta }%
}e_1=0 \\
&  &  &  &  &  &  \\
\eta e_2={{(e_3-e_0)}\over{\sqrt{2}}} &  &
\overline{\eta }e_2=0 &  & \partial _\eta e_2=0 &  &
\partial _{\overline{\eta }}e_2={{(e_3+e_0
)}\over{\sqrt{2}}} \\
&  &  &  &  &  &  \\
\eta e_3={{e_1}\over{\sqrt{2}}} &  & \overline{\eta }%
e_3={{e_2}\over{\sqrt{2}}} &  & \partial _\eta
e_3={{e_2}\over{\sqrt{2}}} &  & \partial _{\overline{\eta
}}e_3=-{{e_1}\over{\sqrt{2}}}\;.
\end{array}
\label{tabla}
\end{equation}
Some explicit results obtained in this way are, for example,
\begin{equation}
Tr_{\{\eta ,\overline{\eta }\}}(I)=Tr_{\{\eta ,\overline{\eta }\}}(\theta
)=Tr_{\{\eta ,\overline{\eta }\}}(\overline{\theta })=0
\;\;,Tr_{\{\eta ,\overline{\eta }\}}(\overline{\theta }\theta )=q \;.
\label{trazas}
\end{equation}
The ``kinetic'' term of the Lagrangian (\ref{sslagr}) is,
\begin{eqnarray}
Tr_{\{t,\eta ,\overline{\eta }\}}[D_{\overline{\theta }}\Phi D_\theta \Phi ]
&=&Tr_{\{t,\eta ,\overline{\eta }\}}\left[\overline{\psi }(t)\psi (t)-\overline{%
\psi }(t)\overline{\theta }\theta T\psi (t)+q\overline{\psi }(t)\overline{%
\theta }\theta \psi (t)T  \right. \nonumber \\
&&\left.\qquad \qquad -\overline{\psi }(t)T\overline{\theta }\theta \psi (t)+qT%
\overline{\psi }(t)\overline{\theta }\theta \psi (t)\right]  \nonumber \\
&&\qquad \qquad +q^2 \int_0^T dt[\stackrel{.}{x}(t)^2+d(t)^2]  \;.
\label{kinacc}
\end{eqnarray}
The explicit calculation of the traces yields the following results,
\begin{equation}
\begin{array}{l}
Tr_{\{\eta ,\overline{\eta }\}}[\overline{\psi }(t)\psi (t)]={2\over{\sqrt{q%
}}}\sinh (\alpha /2)\overline{\chi }(t)\chi (t) \\
\\
Tr_{\{\eta ,\overline{\eta }\}}[\overline{\psi }(t)\overline{\theta }\theta
T\psi (t)]=q\overline{\chi }(t)T\chi (t) \\
\\
Tr_{\{\eta ,\overline{\eta }\}}[\overline{\psi }(t)\overline{\theta }\theta
\psi (t)T]=\overline{\chi }(t)\chi (t)T \\
\\
Tr_{\{\eta ,\overline{\eta }\}}[T\overline{\psi }(t)\overline{\theta }\theta
\psi (t)]=T\overline{\chi }(t)\chi (t)
\end{array}  \;.
\label{xitrazas}
\end{equation}
The first trace is of central importance. In the undeformed case, the
term $\overline{\psi }(t)\psi (t)$ does not contribute. Thus we get
the following free Lagrangian,
\begin{equation}
Tr_{\{t,\eta ,\overline{\eta }\}}[\frac 12D_{\overline{\theta }}\Phi
D_\theta \Phi ]=\int_0^Tdt\{\frac 12q^2[\stackrel{.}{x}(t)^2+d(t)^2]+%
\overline{\chi }(t)[iq\partial _t+m(q)]\chi (t)\}  \label{kinlag}
\end{equation} \,
where the mass parameter $m(q)$ is given by,
\begin{equation}
m(q)=\frac 1{\sqrt{q}}\sinh (\frac \alpha 2)  \;.
\label{masa}
\end{equation}
We take the potential $V(\Phi )$ to be a power series in $\Phi ,$
\begin{equation}
V(\Phi )=\sum_mv_m\Phi ^m  \label{poten} \;.
\end{equation}
It is easy to see that, because of trace over $\{\eta ,\overline{\eta }\},$
the effective contribution of this term amounts to,
\begin{equation}
Tr_{\{t,\eta ,\overline{\eta }\}}[V(\Phi )]=\int_0^Tdt\{q\frac{\partial V(x)%
}{\partial x}d(t)+\frac{\partial ^2V(x)}{\partial x^2}\overline{\chi }%
(t)\chi (t)\}  \;.
\label{poten2}
\end{equation}
Therefore the total action $S[x,\overline{\chi },\chi ]$ is,
\begin{eqnarray}
S[x,d,\overline{\chi },\chi ] &=&
\int_0^T dt \{\frac{1}{2} q^2[\stackrel{.}{x}
(t)^2+d(t)^2]-q\frac{\partial V(x)}{\partial x}d(t)  \nonumber\\
& \; &\;\;\;\;\;\;\;\;\;+ {\overline{\chi}}(t) [i q \partial_t +m(q)-
\frac{\partial^{2} V(x)}{\partial x^2}
]\chi(t) \}  \;.
\label{fulac}
\end{eqnarray}
In order to evaluate the contribution of the fermionic sector, it is
convenient to express $\chi (t)$ in term of the eigenfunctions of
the operator $[iq\partial _t+m(q)-\frac{1}{q}V^{\prime \prime }(x)]$, i.e.,
\begin{eqnarray}
\chi (t) &=&\sum_n\chi _n\varphi _n(t)  \nonumber  \;,\\
\overline{\chi }(t) &=&\sum_n\overline{\chi }_n\overline{\varphi }_n(t)
\nonumber \;,
\label{expanx}
\end{eqnarray}
where,
\begin{equation}
\begin{array}{l}
\chi _n^2=\overline{\chi }_n^2=0 \;,\\
\\
\chi _m\overline{\chi }_n+{ 1\over q}\overline{\chi }_n\chi _m=0 \;,
\end{array}
\label{xn}
\end{equation}
and,
\begin{equation}
\lbrack iq\partial _t+m(q)-V^{\prime \prime }(x)]\varphi _n(t)=\lambda
_n\varphi _n(t)  \;,
\label{eigen}
\end{equation}
with periodic boundary conditions, imposed by supersymmetry. In fact,
requiring periodic boundary conditions for the bosonic component $x(t)$ and
its $q$-supersymmetric variation $\delta x(t),$ given in (\ref{fsstrans}),
one may constrain the boundary condition on $\psi (t)$. Then, by means of
the representation (\ref{psi}), one gets periodic boundary conditions
for $\chi(t).$ Obviously, the same holds for its
$q$-Grassmannian conjugated $\overline{\chi }(t).$
So, the eigenfunctions and eigenvalues are,
\begin{equation}
\begin{array}{l}
\varphi_n(t)=C \, e^{\frac{-i}{q}
\int_0^t dt^{\prime }[\lambda_n -m(q)+ V^{\prime\prime} (x)]} \\
\\
\lambda_n={{2n\pi q} \over {T}}+m(q)
-{1\over T}\int_0^T V^{\prime \prime }(x) \qquad \;;n \in {Z \!\!\! Z}.
\end{array}
\label{autoval}
\end{equation}
Therefore we get the following fermiomic action,
\begin{eqnarray}
S_F [x,\overline{\chi }_n,\chi _n]&\,=\,&
\int_0^Tdt\quad \left\{\overline{\chi }(t)\left[iq\partial _t+m(q)-\frac{\partial
^2V(x)}{\partial x^2}\right]\chi (t)\right\}   \nonumber\\
&\,=\,&\sum_n\lambda _n\overline{\chi }_n\chi _n
\;.
\label{acef}
\end{eqnarray}
Since $d(t)$ is an auxiliary variable we can eliminate it using its
equation of motion. Thus, the action becomes,
\begin{equation}
S[x,\overline{\chi }_n,\chi _n]=\frac 12\int_0^Tdt\quad \{q^2\stackrel{.}{x}%
(t)^2-[V^{\prime }(x)]^2\}+\sum_n\lambda _n\overline{\chi }_n\chi _n
\;.\label{fulacc}
\end{equation}
For the evaluation of the fermionic contribution it is convenient to express the
q-Grassmanian components $\overline{\chi }_n$ and $\chi _n$ in terms of
standard Grassmann variables,
\begin{equation}
\begin{array}{l}
\chi _n=\sigma _ne^{-\frac 12\alpha \sum_k[\overline{\sigma }_k\partial _{%
\overline{\sigma }_k}-\partial _{\sigma _k}\sigma _k]} \\
\\
\overline{\chi }_n=e^{\frac 12\alpha \sum_k[\overline{\sigma }_k\partial _{%
\overline{\sigma }_k}-\partial _{\sigma _k}\sigma _k]}\overline{\sigma }_n
\end{array} \;,
\label{repx}
\end{equation}
where, $\sigma _n^2=\overline{\sigma }_n^2=0$ and $\{\sigma _m,\overline{%
\sigma }_n\}=0.$ The generating functional for this action is,
\begin{equation}
Z=\frac 1N\int Dx\quad e^{\frac i{2\hbar }\int_0^Tdt\{q^2\stackrel{.}{x}%
(t)^2-[V^{\prime }(x)]^2\}}\quad Tr_{\{\sigma ,\overline{\sigma }%
\}}[e^{\frac i\hbar \sum_n\lambda _n\overline{\chi }_n\chi _n}]\;.
\label{partfunction}
\end{equation}
To obtain an effective theory, we ``integrate out" the fermionic fields, i.e,
we evaluate the fermionic contribution to the generating functional,
\begin{equation}
Z_{F} \,=\,
Tr_{\{\sigma ,\overline{\sigma }\}}[e^{\frac i\hbar \sum_n\lambda _n%
\overline{\chi }_n\chi _n}] \;.
 \label{fermtrace}
\end{equation}
We regularize this expression by letting $n$ running from $-N$ to $N$.
Noting that,
\[
\overline{\chi }_n\chi _n=e^\alpha \; \overline{\sigma }_n\sigma _n
\;,\]
the trace is now easily evaluated, leading to,
\begin{equation}
Tr_{\{\sigma ,\overline{\sigma }\}}[e^{\frac i\hbar \sum_n\lambda _n%
\overline{\chi }_n\chi _n}]=(i/\hbar )^{2N+1}\; e^{(2N+1)\alpha }\;
\lambda _{-N}...\lambda _N  \;.
\label{detgrass}
\end{equation}
Since we are mainly interested in the construction of an effective action,
the relevant quantity will be the logarithm of the above expression.
Following ref. \cite{gozzi}, we get
\begin{equation}
Tr_{\{\sigma ,\overline{\sigma }\}}[e^{\frac i\hbar \sum_n\lambda _n%
\overline{\chi }_n\chi _n}]=\sinh \left[ \frac i{2q}\int_0^T\left(
m(q)-V^{\prime \prime }(x)\right) dt\right]\;,
\end{equation}
leading to a generating functional of the form,
\begin{equation}
Z=Z_{+}-Z_{-}  \;,
\label{zeff}
\end{equation}
where,
\[
Z_{\pm}=\frac 1N\int Dx\quad e^{\frac i\hbar \int_0^Tdt\left[ \frac 12q^2%
\stackrel{.}{x}^2(t)-\frac 12[V^{\prime }(x)]^2 \pm \frac 1{2q}m(q)\mp\frac
1{2q}V^{\prime \prime }(x)\right] }\;.
\]

As shown in ref. \cite{cooper} and \cite{gozzi}, this effective generating
functional corresponds to the Hamiltonian,
\begin{equation}
\widetilde{H}_q=\left[
\begin{array}{ccc}
H_q^{+} &  & 0 \\
&  &  \\
0 &  & H_q^{-}
\end{array}
\right]\;,   \label{ham}
\end{equation}
with,
\begin{eqnarray}
H_q^{\pm} &=&\frac 1{2q^2}p^2(t)+\frac 12[V^{\prime }(x)]^2\mp\frac \hbar
{2q}V^{\prime \prime }(x)\pm\frac \hbar {2q}m(q)  \label{h-} \\
&=&\frac {1}{2}\left( \frac{\mp i}{q}p(x)+V^{\prime }(x)\right) \left( \frac
{\pm i}{q}p(x)+V^{\prime }(x)\right) +\frac \hbar {2q}m(q)  \nonumber\;.
\end{eqnarray}
Observe that $\left[ H_q^{+}-\frac 1{2q}m(q)\right] $ and $\left[
H_q^{-}+\frac 1{2q}m(q)\right] $ are positive-semidefined operators, hence
their minimun eigenvalue is $0.$ The corresponding eigenfunctions are,
\begin{equation}
\begin{array}{l}
\varphi _0^{\pm}(x)=e^{\mp qV(x)} \;,
\end{array}
\label{fundstate}
\end{equation}
such that,
\begin{equation}
\begin{array}{l}
H_q^{\pm}\varphi _0^{+}(x)=\mp\frac 1{2q}m(q)\varphi _0^{\pm}(x) \;.
\end{array}
\label{minenergia}
\end{equation}
So, we can see that the presence of the mass term shift the energy of the
ground state away from zero. Moroever, depending whether $q<1,$ the minimun
energy is,
\begin{equation}
E_o=\frac 1{2q}m(q) \;, \label{geq<1}
\end{equation}
belonging to the eigenstate,
\begin{equation}
\varphi _0=\left[
\begin{array}{c}
e^{-qV(x)} \\
0
\end{array}
\right]   \label{gsq<1}
\end{equation}
or $q>1,$ with eigenvalue,
\begin{equation}
E_o=-\frac 1{2q}m(q)  \label{geq>1}
\end{equation}
and eigenstate
\begin{equation}
\varphi _0=\left[
\begin{array}{c}
e^{-qV(x)} \\
0
\end{array}
\right]   \label{gsq>1}
\end{equation}
Observe that in any case the energy of the ground state is less than zero,
thus supersymmetry is explicitely broken.

We end up with some concluding remarks. Starting from a non commutative
set of coordinates we considered symmetry operations on them, we dealt
with fields over such coordinates, we represent them on certain functional
spaces and we took traces  as a replacement of integration. Such a procedure
is certainly not unique. The selection of a representation is essential and
we have no criteria for such a choice. One possibility is to think of such
a selection on the same footing as the selection of certain dynamical system.

Regarding the breaking of supersymmetry, it is important to realize that,
although we finnally get an explicit breaking, at the starting operatorial
level we have the full power of supersymmetry to restrict the possible
operatorial Lagrangians we can write, in this respect it is similar to
a spontaneous breaking.


\begin{thebibliography}{1}
\bibitem[1]{con}  A. Connes, {\it{Noncommutative Geometry}},
(Academic Press, London, 1994).

\bibitem[2]{conl}  A. Connes and J. Lott, Nucl. Phys. B 18,
Suppl. 29-47(1990),(1991).


\bibitem[3]{man}  Y. I. Manin, {\it{Quantum group and non-commutative
geometry}}, preprint Montreal Univ. CRM-1561 (1988).


\bibitem[4]{drinf}  V. G. Drinfeld, {\it{Quantum Groups, }}Proc. Int.
Congr. Math. Berkeley 1986, vol. 1, 798.

\bibitem[5]{fad}  L.D. Faddeev, {\it{Integrable models in
(1+1)-dimensional quantum field theory,}} Lectures in Les Houches 1982,
Elsiever Science Publishers B.V., 1984.


\bibitem[6]{wit}  E. Witten, Nucl. Phys. B185 (1981), 513.

\bibitem[7]{cooper}  F. Cooper and B. Freedman, Ann. of Phys. (NY) 146
(1983), 262.


\bibitem[8]{gozzi}  E. Gozzi, Phys. Rev. D 28 (1983), 1922.

\end{thebibliography}
\end{document}